# A "solid" approach to biopharmaceutical stabilisation.

Aswin Doekhie[1], Rajeev Dattani[2], Yun-Chu Chen[1], Yi Yang[3], Andrew Smith[4], Alex P. Silve[1], Francoise Koumanov[5], Karen J. Edler[1], *Kevin J. Marchbank[3], *Jean van den Elsen[6] and *Asel Sartbaeva[1].

[1]Department of Chemistry, [2]European Synchrotron Research Facility, [3]Institute of Cellular Medicine, [4]Diamond Light Source, [5]Department for Health, [6]Department of Biology and Biochemistry. *share senior authorship.

[1,5,6] University of Bath, Claverton Down, Bath, BA2 7AY, United Kingdom.

[2]ESRF, 71 avenue des Martyrs, CS 40220, 38043 Grenoble Cedex 9, France.

[3]Newcastle University, Medical School, Newcastle upon Tyne, NE2 4HH, United Kingdom.

[4]Diamond Light Source Ltd, Harwell Campus, Didcot, OX11 0DE, United Kingdom.

**Summary**

**Ensilication is a technology we developed that can physically stabilise proteins in silica without use of a pre-formed particle matrix[1]. Stabilisation is done by tailor fitting individual proteins with a silica coat using a modified sol-gel process[2]. Biopharmaceuticals, e.g. liquid-formulated vaccines with adjuvants, have poor thermal stability. Heating and/or freezing impairs their potency [3-5]. As a result, there is an increase in the prevalence of vaccine-preventable diseases in low-income countries even when there are means to combat them [6-8]. One of the root causes lies in the problematic vaccine 'cold-chain' distribution[3,4,9-12]. We believe that ensilication can improve vaccine availability by enabling transportation without refrigeration. Here, we show that ensilication stabilises tetanus toxoid C fragment [13] (TTCF) and demonstrate that this material can be stored and transported at ambient temperature without compromising the immunogenic properties of TTCF *in vivo*. TTCF is a component of the diphtheria, tetanus and pertussis (DTP) vaccine. To further our understanding of the ensilication process, and its protective effect on proteins we have studied the formation of TTCF-silica nanoparticles via time-resolved Small Angle X-ray Scattering [14] (SAXS). Our results reveal ensilication to be a staged diffusion-limited cluster aggregation (DLCA) type reaction, induced by the presence of TTCF protein at neutral pH. Analysis of scattering data indicates tailor fitting of TTCF protein. The experimental *in vivo* immunisation data**

**confirms the retention of immunogenicity after release from silica. Our results suggest that we could utilise this technology for multicomponent vaccines, therapeutics or other biopharmaceuticals that are not compatible with lyophilisation.**

Biopharmaceuticals are biologically derived active compounds directed for therapeutic or diagnostic usage. These can be identified as, for example, monoclonal antibodies, vaccines, cells and blood components. Many of these are proteins and not stable *ex vivo*. They must be stabilised to increase their shelf-life. The main approach in stabilisation of biopharmaceuticals is regulating temperature. Maintaining low (2-8 °C) temperatures retains native states of these compounds by lowering the kinetic energy. For many of the protein-based biopharmaceuticals, freezing is not beneficial as this weakens hydrophobic interactions, inherent in proteins, which may refold in a different conformation upon thawing. Within cold-chain transportation, from manufacturing to endpoint destination, temperature control has proven challenging. Fluctuations in temperature, heating/freezing[3], can affect biopharmaceuticals resulting in aggregation, protein unfolding leading towards loss of potency. Our previous research describes a new alternative to stabilise biopharmaceuticals at ambient temperatures using sol-gel technology [1]. Sol-gel is based upon the polymerisation of a monomeric species that can undergo condensation with another monomer of its species [2]. This eventually forms a particle as product which, in contact with other particles forms gels. There are many compounds that are compatible with this system, however our method uses a silica precursor, tetraethyl orthosilicate (TEOS). Hydrolysis of TEOS results in orthosilicic acid [15,16]. This is a central Si atom bound with four hydroxyl (OH-) groups in a tetrahedral conformation and allows for rapid polymerisation under optimised conditions. The benefit here is the negative charge of silicic acid which we employ via electrostatics to direct it towards the positively charged biologically active component. Therefore, we can coat silica around a biologically active compound. The protein loaded silica powder obtained is resilient against fluctuations in temperature. TTCF is a ~52 kilodalton (kDa) fragment of the full tetanus neurotoxin and is a potent immunogenic protein. The DTP vaccine, which includes TTCF, has been shown to be susceptible to problems within cold chain transportation. This is represented by 16% lower vaccination coverage for DTP[8] in countries with poor infrastructure and results in higher prevalence of disease. Using TTCF as a model we demonstrate that ensilication could be an ideal solution to overcome the challenges of biopharmaceutical 'cold-chain' transportation.

We have previously described the retention of TTCF protein structure via ensilication at primary and tertiary level, including heat treated material, using biochemical methods [1]. In the present study, we looked at two opposite experimental resolutions. Time resolved (*in situ*) SAXS was employed to assess tailor fitting of TTCF in silica. This is a powerful tool that can assess morphology of proteins and particles formed at nanometre scale over time. It provides the average shape and size of (*in situ*) particle growth using the radius of gyration ($R_g$) and allows for fitting of mathematical models to elucidate protein-silica particle formation [14,17]. This could observe and possibly confirm tailored silica coating of TTCF at molecular level. On the other side of the spectrum, *in vivo* immunisation experiments could confirm feasibility of the end stage application for our methodology creating an opportunity to solve real-world challenges.

Initial assessment of the protein scattering data confirmed a monodisperse solution of native TTCF (Supplementary S1, S2). Calculated parameters elucidate TTCF to be a prolate ellipsoid. The two independent SAXS experiments performed were intended to provide resolution of the onset and longer evolution of TTCF ensilication.

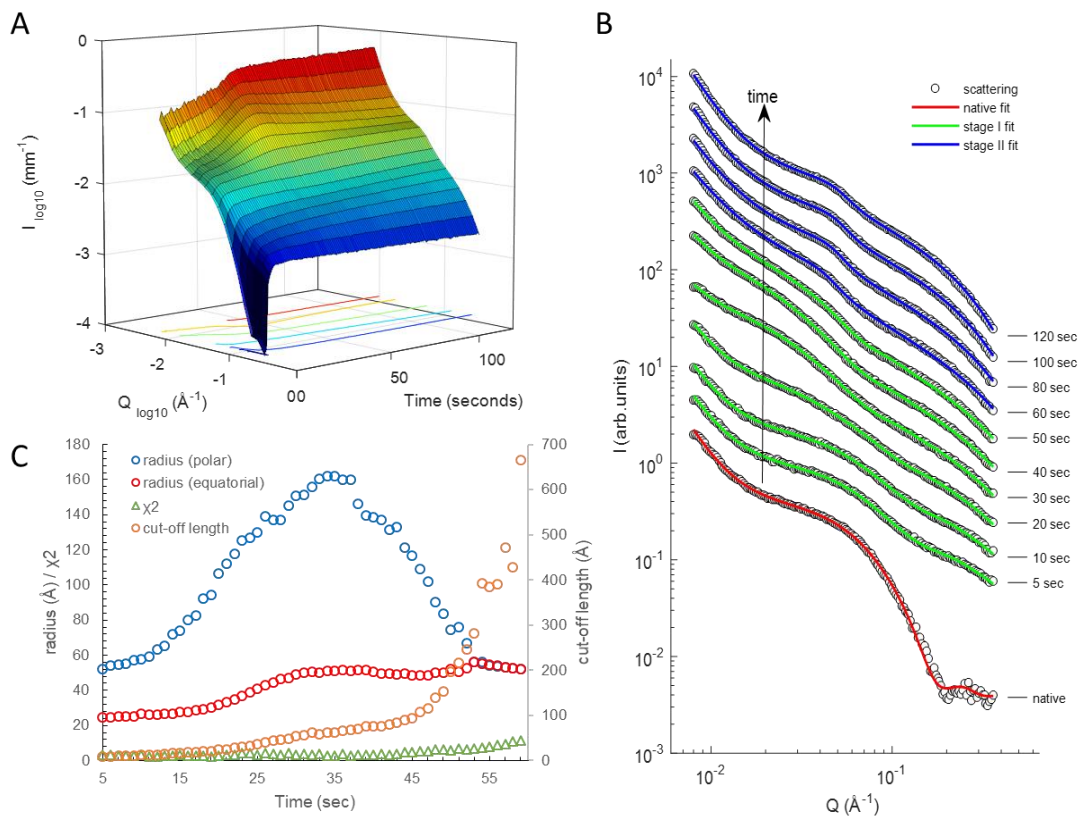

**Figure 1. SAXS scattering of continuous flow TTCF ensilication.** (A) 3D plot of the first 120 seconds during the ensilication of TTCF. Observed is the development of silica particle growth upon onset of polymerisation. (B) Stacked plot of scattering for multiple time points during the experiment. Scattering scaled 1 log for each chosen time point. Red, green and blue fittings use a power law + ellipsoid, power law + ellipsoid + mass fractal and broad peak + mass fractal model respectively (C) Batch analysis data fit results of acquired SAXS signal. Cut-off length (Å) = static length of silica particles coating TTCF, $\chi^2$ = residuals of fit for stage I fit model (Supplementary).

The use of a flow-cell setup (figure S5) with sample agitation (Diamond Light Source, i22) allowed continuous measurement of the onset of TTCF ensilication. The $q$-range (figure 1, A) at $0.008 \leq q \leq 0.35$ Å$^{-1}$ displayed the rapid development of protein-silica particles after addition of hydrolysed TEOS. There is a change in scattering (figure 1, B) that is observed at high $q$ ($0.20 \leq q \leq 0.35$ Å$^{-1}$) and there are fluctuations in the low $q$-range ($0.008 \leq q \leq 0.04$ Å$^{-1}$). The high $q$ signal appears immediately after the hydrolysed TEOS is added which complements with other studies[18] and indicates silica condensation induced by the presence of charged residues of the protein[19]. This is indicative of silica coating of TTCF via ensilication and after approximately 40 seconds the protein signal has shifted towards the low $q$ range therefore providing evidence of particle growth caused by ensilication. Supporting this observation is the fractal structure formed at high $q$. After 60 seconds protein scattering gradually decreases and an amorphous peak is formed, situated around the identical $q$ where the apex of native TTCF is present.

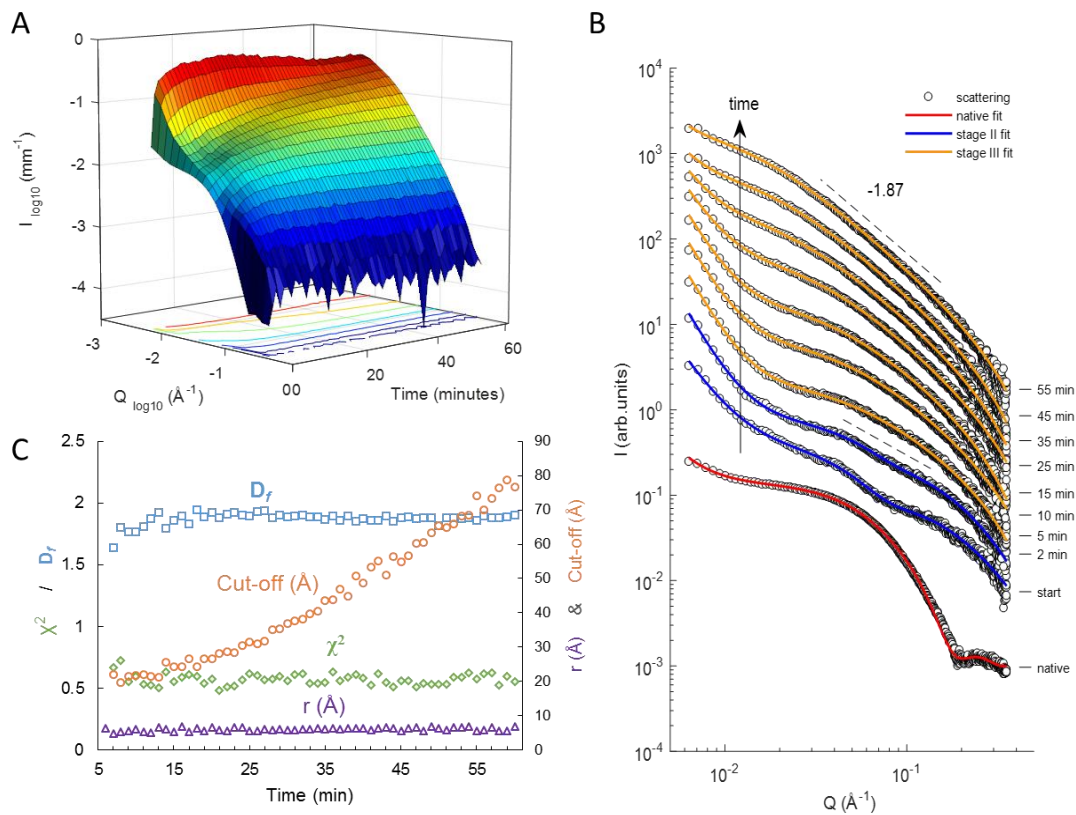

**Figure 2. SAXS scattering of TTCF ensilication over time.** (A) 3D plot of the 60 minutes during the ensilication of TTCF. Observed is the development of silica particle growth. (B) Stacked plot of scattering for multiple time points during the experiment. Scattering scaled for clarity with added fits. Truncated high $q$ for clarity. Red, blue and orange fittings use a power law + ellipsoid, broad peak + mass fractal, power law + mass fractal model respectively. (C) Batch analysis data on mass fractal fitting of acquired SAXS signal. $D_f$ = fractal dimension, r (Å) = fractal radius, cut-off length (Å) = static length of aggregating silica particles, $\chi^2$ = residuals of fit.

Time resolved SAXS (ESRF, ID02) of non-agitated capillary setup (figure S6) TTCF ensilication displayed similar behaviour at $0.006 \leq q \leq 0.51$ (Å$^{-1}$) (figure 2, A). Here, the ensilication process was monitored after *ex situ* initiation of TTCF ensilication. After measurement was started, the amorphous peak at *q*-range of $0.03 \leq q \leq 0.08$ (Å$^{-1}$) (figure 1 and 2, B) was identified in both data sets. Additionally, the rapid increase in high-*q* signal is present here as well. Observed is the extension of this slope from $0.03 \leq q \leq 0.1$ (Å$^{-1}$) and indicated mass fractal growth of silica gel around TTCF ensilicated silica particles forming larger aggregates. There are visible differences in scattering between agitated and non-agitated sample that suggested this reaction to be diffusion controlled (figure S3). Fitting done on both data sets (figure 1&2, C) using various models indicate three stages during ensilication (figure 3).

<u>Stage I</u>: Formation and growth of low *q* particle radii initiated at the onset of ensilication (figure 1 A, until ~40 sec). Here, ensilication is observed via the increase in ellipsoidal radii (polar, equatorial) over time, indicating silica forming around the protein.

<u>Stage II</u>: Broad amorphous peak formation due to aggregation of silica.

<u>Stage III</u>: Fractal growth of silica particles associated with aggregation which relates to increase in fractal dimension and cut-off length (figure 1, C). This is confirmed by the mass fractal dimension ($D_f$) during the longer evolution (figure 2, C) which provides an increase to $D_f = 1.87$ and has been shown to be indicative of fast silica growth and defines ensilication as a diffusion limited cluster aggregation (DLCA) type process[20] (figure 3).

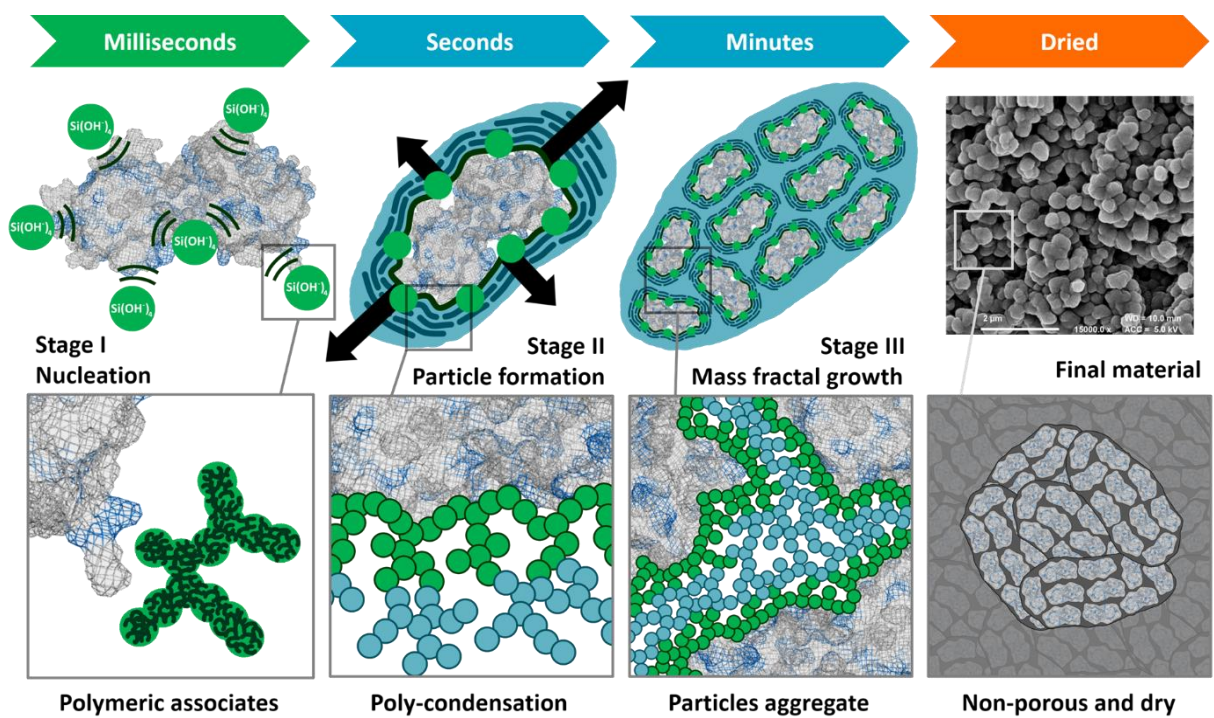

**Figure 3. Graphical representation of TTCF ensilication.** Nucleation, induced via electrostatics, initiates ensilication at positive external residues (lysine, arginine, blue) present on the protein. Poly-condensation of SiOH$_4$ results in silica coating of individual proteins (stage I, green) which triggers aggregation (stage II, blue). Vacuum filtration of the then turbid solution results in dried powder material containing protein loaded silica nanoparticles (FE-SEM at 15,000x magnification).

Our findings are further supported by SAXS carried out a lower q-range (0.0008 to 0.05 A$^{-1}$) displaying large particle formation after silica addition (figure S4). Particle sizes increasing to a stable range of 2000 - 2200 Å (200-220 nm) are observed. This data suggests that once TTCF is coated with silica, it is prone to aggregate and form a stable complex. FE-SEM imaging of the final dried material confirm these sizes (figure 3). Overall, the SAXS data provides evidence of a fast DLCA process where, once TTCF is coated, on average 2000 Å aggregates of protein-silica particles will form.

To confirm our protective capacity *in vivo*, we immunised mice with TTCF or with TTCF after ensilication and obtained serum samples. Enzyme Linked Immuno-Sorbent Assay [21] (ELISA) using native TTCF coated plates were employed to analyse the resultant immune response. As expected from previous studies[22], we observed an early stage immune response in the mice injected with native TTCF and this immune response was matched when the mice were injected with TTCF which had been stored as ensilicated powder and released prior to immunisation (figure 4).

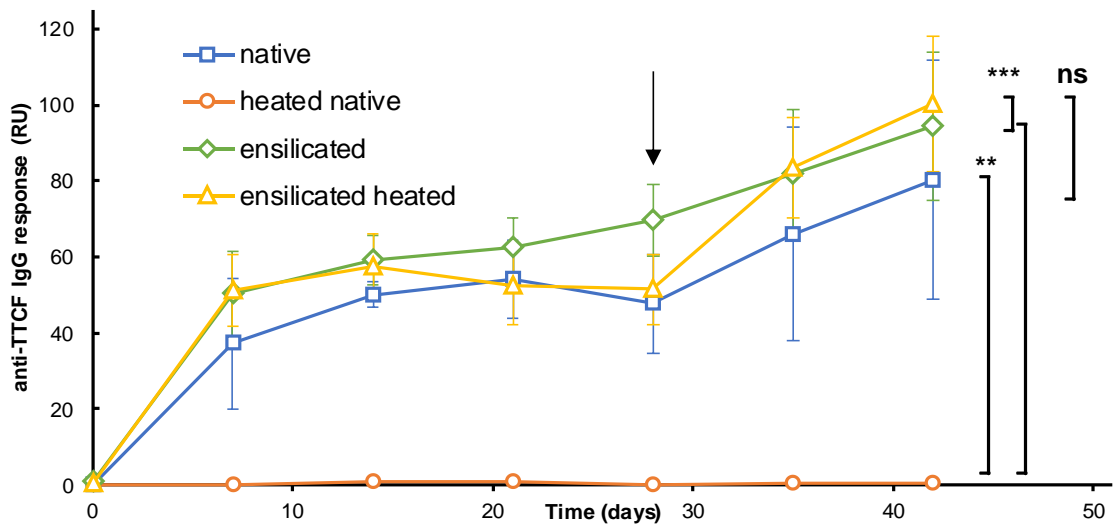

**Figure 4**. **Serum IgG response from mice immunised with TTCF after ensilication.** Normalised absorbance values against monoclonal 10G5 (RU: relative units). Sample groups consisted of 5 mice which were immunised with 5 µg of TTCF at the start of experiment and boosted at 28 days (arrow). Data are expressed as mean +/- SD (n=5). Statistical analysis was performed applying one-way ANOVA with post-hoc Tukey HSD test. Comparison vs native TCCF (ns = nonsignificant) or vs native TTCF heated to 80 °C to denature and inactivate ((* $p \leq 0.05$, ** 0.01 and *** 0.001,)

The results confirmed the protective capacity of ensilication to maintain normal protein/epitope conformation at ambient temperature, as well as when the ensilicated material was heated at 80 °C for 2 hours. The ensilicated material used during this study was stored in powdered form for 1 month at room temperature and then transported using commercially available means without any specialised equipment. The ensilicated TTCF was protected from elevated temperature denaturation, as demonstrated by the lack of specific immune response when unprotected and heat-denatured TTCF was injected into mice (Figure 4 open circles). This demonstrates the ease and efficacy of use of our methodology. Additionally, flow cytometry analysis of day 42 (post-immunisation) splenocytes (figure S7) suggested ensilication of TTCF to have no influence on the measured immune response.

We believe that this new methodology has given rise to another solution for biopharmaceutical stabilisation. In this study, we have confirmed ensilication using SAXS and demonstrated the protective capacity of ensilication via *in vivo* analysis, i.e. the immunogenicity for intact TTCF is maintained. All biopharmaceuticals have unique functions that require different environments for their operations. Therefore, for some, lyophilisation might work better as a means of stabilising and transporting the bioactive compounds. We propose that those which are not compatible with lyophilisation could be remediated via ensilication.

**Acknowledgements**

This work benefited from the use of the SasView application, originally developed under NSF award DMR-0520547. SasView contains code developed with funding from the European Union's Horizon 2020 research and innovation programme under the SINE2020 project, grant agreement No 654000. Diamond Light Source beamtime at i22 was granted on project SM-14148. ESRF beamtime at ID02 was granted on project SC4282. AD thanks the Annett Trust and the University of Bath for funding. This work was supported in part by the BBSRC (JvdE, KJM; BB/N022165/1), Newcastle Universities Confidence in Concept scheme (KJM) and MRC (FK: MR/P002927/1). The authors also thank the assistance of the staff in the comparative biology centre at Newcastle University. AS thanks the University of Bath and the Royal Society for funding. Our gratitude goes out to George Agbakoba for flow cytometry staining and initial analysis. We thank Ursula Potter for FE-SEM imaging of the final material.

**Author Contributions**

A. Sartbaeva designed the study. A. Sartbaeva, F.K., J.vd.E. and K.J.M conceived the experiments. A.D. prepared TTCF for experiments. A.D., YC.C., A. Sartbaeva and A. Smith performed the experiment at i22, Diamond, with support from F.K. and K.J.E.

A.D., YC.C., A. Sartbaeva and R.D. performed the experiment at ID02, ESRF. SAXS data analysis was performed by A.D. with input and discussion from R.D., A.P.S., K.J.E., F.K., J.vd.E. and A. Sartbaeva. R.D. and A.P.S., provided computational assistance.

K.J.M. and Y.Y. conducted the *in vivo* experiments. A.D. ran the serum analysis and statistics with input and discussions from F.K., Y.Y., J.vd.E. and K.J.M.

K.J.M. analysed the flow cytometry data. A.D. wrote the manuscript with contributions from all co-authors.